\documentclass[aps,prb,twocolumn,floatfix,showpacs]{revtex4}
\usepackage{amsmath,amssymb,graphicx,bm}
\begin{document}
\title{Kondo behavior in the asymmetric Anderson model: Analytic
  approach}

\author{V.  Jani\v{s}} \author{P. Augustinsk\'y}

\affiliation{Institute of Physics, Academy of Sciences of the Czech
  Republic, Na Slovance 2, CZ-18221 Praha 8, Czech Republic}
\email{janis@fzu.cz, august@fzu.cz}

\date{\today}

\begin{abstract}
  The low-temperature behavior of the asymmetric single-impurity
  Anderson model is studied by diagrammatic methods resulting in
  analytically controllable approximations. We first discuss the ways
  one can simplify parquet equations in critical regions of
  singularities in the two-particle vertex. The scale vanishing at the
  critical point defines the Kondo temperature at which the
  electron-hole correlation function saturates. We show that the Kondo
  temperature exists at any filling of the impurity level. A
  quasiparticle resonance peak in the spectral function, however,
  forms only in almost electron-hole symmetric situations. We relate
  the Kondo temperature with the width of the resonance peak. Finally we
  discuss the existence of satellite Hubbard bands in the spectral
  function.
\end{abstract}

\maketitle 

\section{Introduction}\label{sec:Intro}
Single-impurity Anderson model (SIAM) has attracted much interest of
theoretical physicists from its introduction in the early sixties of
last century as a simple model for formation of local magnetic
moments.\cite{Anderson61} One of the reasons for the interest in SIAM
has been its connection with the $s-d$ exchange model and the Kondo
effect \cite{Kondo64} demonstrating a nontrivial strong-coupling
behavior. The model won on importance after finding an exact solution
with an algebraic Bethe ansatz.\cite{Tsvelik83} Revival of interests
in reliable methods suitable for solving impurity models was brought
with the concept of the dynamical mean-field theory of correlated
lattice electrons.\cite{Georges96} In particular, methods addressing
dynamical properties of impurity models have been demanded in the
mean-field description of strongly correlated systems. Since the exact
algebraic approaches cover only static properties, a number of
approximate analytic and numerical approaches have been revisited in
the effort to provide an accurate impurity solver for strong electron
correlations.

The direct way to treat the dynamics of SIAM is to use the
perturbation theory in the interaction strength. It is known that such
a perturbation series converges for all interaction strengths and
already second order gives rather accurate
results.\cite{Yosida70,Zlatic83} Unrenormalized weak-coupling
expansions break down in critical regions of any phase transition
signaled by a singularity in a two-particle correlation function.
Hence, extension of perturbative expansions to the strong-coupling
regime of translationally invariant lattice models is questionable.
On the other hand, strong-coupling expansions based on the
infinite-interaction model \cite{Keiter70,Pruschke89} fail to
reproduce the Fermi liquid regime in the weak-coupling limit. A
dynamical theory reliably describing the transition from the
weak-coupling Fermi liquid to a strong-coupling solution is still
missing.

Presently, the most comprehensive quantitative approaches producing
dynamical properties of the impurity models are quantum Monte Carlo
(QMC)\cite{Fye88} and the numerical renormalization group
(NRG).\cite{Wilson75} The latter scheme was first devised by Wilson
when solving the $s-d$ exchange model and later extended to SIAM in
Ref.~\onlinecite{Krishnamurthy80}. Although designed for static
thermodynamic properties, NRG can be used to some extent for the
calculation of dynamical quantities as well.\cite{Costi94, Bulla98}
Since the quantum Monte Carlo is restricted to rather high
temperatures, NRG represents now-a-days the most accurate quantitative
low-temperature and low-energy solution of impurity problems for
intermediate couplings.\cite{Hewson93}

Unbiased numerical schemes such as NRG and QMC are reliable only for
nonsingular quantities. A critical behavior can be deduced in these
schemes only from extrapolations. When no simple scaling law holds we
cannot make definite conclusions on the existence and character of the
critical behavior in the strong-coupling regime of correlated
electrons only from numerical data. An analytic theory is needed for
this purpose. We recently developed an analytic approximate scheme
summing selected classes of Feynman diagrams for the two-particle
vertex.\cite{Janis07a} The approximation is justified in the critical
region of the Bethe-Salpeter equation in the electron-hole scattering
channel and uses a simplified version of the parquet equations for the
electron-hole and electron-electron irreducible vertex functions. We
demonstrated that in the symmetric case of SIAM, the low-energy Kondo
scale is qualitatively correctly reproduced (universal features
thereof) in this approximation.\cite{Janis07a}

In this paper we discuss the strong-coupling Kondo behavior in SIAM in
a more general context beyond the symmetric case. For this purpose we
use simplified parquet parquet equations. We introduce a general
method of simplification of the parquet equations for impurity models.
It is based on a partial separation of short- and long-range energy
scales induced by a singularity in one of the Bethe-Salpeter equations
for the two-particle vertex. In the critical region of a singularity
in the vertex function we replace regular functions (no critical
long-range fluctuations) by constants and keep dynamical only the
singular ones with critical fluctuations. In this way we do not affect
the universal features of the critical behavior. Unlike the classical
criticality we are, however, unable to uniquely separate short- and
long-range fluctuations in the strong-coupling limit of correlated
electrons. The universal and non-universal properties are mixed up
there. We hence have to decouple the short- and long-range
fluctuations with an additive approximation to make the ansatz of
separation of scales consistent. We can do it in several ways. Here we
discuss two of them. In the first one, introduced in
Ref.~\onlinecite{Janis07a}, we demand fulfilling of the parquet
equations for regular vertices averaged over short-range fluctuations
to obtain constants. This is a mean-field-type approximation that can
be applied in the whole region of the input parameters also beyond the
critical region of the Bethe-Salpeter equation. It is a fully
consistent approximation in the Fermi-liquid regime. Another, more
accurate way of decoupling short- and long-range fluctuations
introduced in this paper is to use systematically the dominant small
energy scale shaping the singularity in the vertex function to reduce
the frequency dependence of regular functions. The latter approach is
more precise as concerns the critical behavior but less flexible as
concerns applicability outside the critical region.

Studying the Kondo behavior in SIAM we distinguish its two different
aspects. First, we introduce a small dimensionless scale connected
with a singularity in the two-particle vertex that we call the Kondo
scale. From this scale we derive the Kondo temperature as a cutoff
screening the actual singularity that cannot be reached in impurity
models. Next we analyze the behavior of one-electron functions in the
Kondo limit, i.~e., with a quasi-divergent two-particle vertex. We
demonstrate the existence of a narrow Kondo resonance peak near the
Fermi energy for small deviations from the symmetric situation. We
calculate the position of the Kondo peak and estimate its width. We
also discuss the existence of the satellite Hubbard bands in the
strong-coupling limit.

It is more elaborate to study asymmetric situations in impurity models
than the symmetric case. We must introduce an effective chemical
potential controlling the occupation of the impurity level. To
guarantee consistency of approximations, the chemical potential must
be determined self-consistently from the actual filling of the
impurity level. It means that independently of the form of
one-electron propagators we use in the approximate treatment we have
to calculate impurity filling from the full spectral function with the
self-energy obtained in the parquet approximation. We hence must
achieve a static one-particle self-consistency. Only in this way we
avoid spurious (Hartree-type) transitions in the strong-coupling
regime.

The paper is organized as follows. In Sec.~\ref{sec:Parquets} we
introduce the parquet equations for impurity models. The
simplifications of the parquet equations in the critical region of a
singularity in the Bethe-Salpeter equations are discussed in
Sec.~\ref{sec:Simplified}. Two-particle functions in the Kondo regime
are calculated in Sec.~\ref{sec:Kondo-scale} and the one-particle
functions in Sec.~\ref{sec:Self-energy}.

\section{Parquet equations for impurity models}\label{sec:Parquets}

We start our investigation with the Hamiltonian of SIAM that reads
\begin{multline}\label{eq:H-SIAM}
  \widehat{H} = \sum_{{\bf k}\sigma} \epsilon({\bf k})
  c^{\dagger}_{{\bf k}\sigma} c^{\phantom{\dagger}}_{{\bf k}\sigma}+
  E_d\sum_\sigma d^{\dagger}_\sigma d_\sigma \\ + \sum_{{\bf
      k}\sigma}\left(V^{\phantom{*}}_{{\bf k}}d^{\dagger}_\sigma
    c^{\phantom{\dagger}}_{{\bf k}\sigma} + V^*_{{\bf k}}
    c^{\dagger}_{{\bf k}\sigma} d^{\phantom{\dagger}}_\sigma\right) +
  U\widehat{n}^d_\uparrow\widehat{n}^d_\downarrow\ .
\end{multline}
We denoted $\widehat{n}^d_\sigma = d^{\dagger}_\sigma d_\sigma$.  When
calculating the grand potential and thermodynamic properties of the
impurity site we can explicitly integrate over the degrees of freedom
of the delocalized electrons. To this purpose we standardly replace
the local part of the propagator of the conduction electrons by a
constant $\Delta(\epsilon) = \pi \sum_{\bf k} |V_{\bf k}|^2
\delta(\epsilon - \epsilon({\bf k})) \doteq \Delta $ the value of
which we set as the energy unit. With this simplification we expand
all dynamical and thermodynamic quantities in powers of the
interaction strength $U$ and rearrange appropriately the resulting
series.
 
The principal idea of the parquet approach is to derive the
self-energy and other thermodynamic and spectral properties via a
two-particle vertex $\Gamma$. This vertex is standardly represented
via Bethe-Salpeter equations.  The Bethe-Salpeter equations express
the full two-particle vertex by means of irreducible ones that can be
generically represented as\cite{Janis99b}
\begin{subequations}\label{eq:BS-parquet}
  \begin{equation}\label{eq:BS-def}
    \Gamma = \Lambda^\alpha + [\Lambda^\alpha G G]_\alpha \star \Gamma
  \end{equation}
  where $\Lambda^\alpha$ is the irreducible vertex in channel
  $\alpha$, the brackets $[\ldots]_\alpha$ stand for an appropriate
  interconnection of internal variables in the chosen scattering
  channel $\alpha$ and $\star$ represents summation over intermediate
  states created during the scattering process. We use in our
  consideration only singlet electron-hole and electron-electron
  scattering channels.

  The consistency of approaches with more two-particle vertices is
  guaranteed by the parquet equation that reads
  \begin{equation}\label{eq:Parquet-def}
    \Gamma = \Lambda^{eh} + \Lambda^{ee}  - \mathcal{I}\ .
  \end{equation}
\end{subequations}
We introduced a vertex $\mathcal{I}$ irreducible in both channels that
in the parquet approximation is replaced by the bare interaction $U$.
The parquet equation~\eqref{eq:Parquet-def} says that the full
two-particle vertex is a sum of the irreducible vertices from the
electron-hole channel ($\Lambda^{eh}$) and from the electron-electron
channel ($\Lambda^{ee}$) from which the completely irreducible vertex
(bare interaction) is subtracted. Equation \eqref{eq:Parquet-def} is
used in each Bethe-Salpeter equation~\eqref{eq:BS-def} to exclude the
full vertex $\Gamma$ and to close the approximation for the
irreducible vertices $\Lambda^{eh}$ and $\Lambda^{ee}$. In the
impurity case the dynamical variables in the vertex functions are
Matsubara frequencies. Hence the explicit form of the parquet
equations with the electron-hole and electron-electron singlet
channels read
\begin{subequations}\label{eq:BS}
  \begin{multline}\label{eq:BS-eh}
    \Lambda^{ee}_{\uparrow\downarrow}(i\omega_n,i\omega_{n'},i\nu_m) =
    \mathcal{I}^{eh}_{\uparrow\downarrow}(i\omega_n, i\omega_{n'},
    i\nu_m)\\ - \frac 1\beta
    \sum_{n''}\Lambda^{eh}_{\uparrow\downarrow}(i\omega_n,
    i\omega_{n"};
    i\nu_m) G_\uparrow(i\omega_{n''})   G_\downarrow(i\omega_{n'' + m})\\
    \times \left[\Lambda^{ee}_{\uparrow\downarrow}(i\omega_{n"},
      i\omega_{n'}; i\nu_m) + \Lambda^{eh}_{\uparrow
        \downarrow}(i\omega_{n"}, i\omega_{n'}; i\nu_m) \right. \\
    \left. - \mathcal{I}_{\uparrow \downarrow}(i\omega_{n"},
      i\omega_{n'} i\nu_m) \right]\ .
  \end{multline}
  and
  \begin{multline}\label{eq:BS-ee}
    \Lambda^{eh}_{\uparrow\downarrow}(i\omega_n, i\omega_{n'}; i\nu_m)
    =
    \mathcal{I}_{\uparrow\downarrow}(i\omega_n, i\omega_{n'}; i\nu_m) \\
    - \frac 1\beta \sum_{n''} \Lambda^{ee}_{\uparrow
      \downarrow}(i\omega_n, i\omega_{n''}; i\nu_{m + n' - n''})
    G_\uparrow(i\omega_{n''})\\ \times G_\downarrow(i\omega_{n + n' +
      m - n''})\left[ \Lambda^{ee}_{\uparrow\downarrow}(i\omega_{n''},
      i\omega_{n'}; i\nu_{m + n - n''}) \right. \\ \left. +
      \Lambda^{eh}_{\uparrow \downarrow} i\omega_{n''}, i\omega_{n'};
      i\nu_{m + n - n''})\right. \\ \left. - \mathcal{I}_{\uparrow
        \downarrow} i\omega_{n''}, i\omega_{n'}; i\nu_{m + n -
        n''})\right] .
  \end{multline}\end{subequations}
These equations form a set of coupled nonlinear integral equations
that cannot be solved by available analytic means. Their numerical
solution with $\mathcal{I} = U$ is available only for high
temperatures\cite{Bickers92} in the region irrelevant for quantum
effects induced by strong electron correlations. Since we do not know
the exact form of the completely irreducible vertex $\mathcal{I}$ we
have to resort in the parquet approach to approximations anyway.

\section{Two-particle criticality and simplified parquet equations}
\label{sec:Simplified}

For most of physically interesting situations it is not necessary to
know the detailed form of the completely irreducible vertex
$\mathcal{I}$. If it is a regular function it is reasonable to replace
it with the bare interaction, which is standardly done in the parquet
approximation. It is then futile to try to solve the parquet equations
with a tremendous effort in regions where no significant deviations
from Fermi liquid or gas are present. The principal area of
application of the parquet approach should be in the low-temperature
strong-coupling region where the standard weak-coupling schemes fail.

The critical region of a singularity in one of the Bethe-Salpeter
equations is the situation where the difference between the parquet
theory and other weak-coupling schemes becomes significant. Parquet
equations due to their nonlinear character introduce a two-particle
self-consistency enabling to handle properly the critical behavior.
In particular, the two-particle self-consistency effectively
suppresses spurious non-physical singularities of one-particle
approximations. That is why we try to solve the parquet equations, at
least in an approximate way, in the critical region of a singularity
in Bethe-Salpeter equations. Fortunately, singularities with divergent
vertex functions offer a natural way of a simplification of the
parquet equations. This simplification is a separation of long-range
fluctuations of divergent quantities from the short-range ones of
regular functions. In the critical region of a singular vertex we then
treat finite differences in regular functions as negligible with
respect to macroscopic changes shaping the singular behavior.
 
To make the simplification of the parquet equations effective we must
rearrange the weak-coupling expansion in such a way that we first
reach the critical region of a singularity in Bethe-Salpeter
equations. In SIAM it is achieved in the random-phase approximation
(RPA), that is, the Bethe-Salpeter equation in the electron-hole
channel with the bare interaction and the Hartree one-electron
propagators. The two-particle vertex in Matsubara frequencies then
reads
\begin{equation}\label{eq:RPA-vertex}
  \Gamma^{RPA}(i\nu_m) = \frac U{1 + U \chi_{eh}(i\nu_m)}
\end{equation}
where we denoted $\chi_{eh}(i\nu_m) = \beta^{-1} \sum_n
G_\uparrow(i\omega_n) G_\downarrow(i\omega_{ n + m})$ the dynamical
electron-hole bubble. It is easy to evaluate the electron-hole bubble
with the Hartree propagators $G(x + iy) = 1/(x -\overline{\mu} + i
\text{sgn}(y)(\Delta + |y|))$ to be
\begin{multline}\label{eq:Chi_eh}
  \chi_{eh}(z) = - \int_{-\infty}^{\infty}\frac{dx}\pi f(x) \left[
    \frac1{x -\overline{\mu} + z + i\Delta}\right. \\ \left.  +
    \frac1{x -\overline{\mu} - z + i\Delta}\right] \frac \Delta {[(x -
    \overline{\mu})^2 + \Delta^2]^2}
\end{multline}
where $\overline{\mu}= E_d +Un$ is an effective chemical potential
measuring deviations from the symmetric case $\overline{\mu} = 0$ and
$f(x) = 1/(1 + \exp\{ \beta x\})$ is the Fermi function. We used $0< n
<1$, the actual occupation per spin of the impurity level.  Since the
static value $\chi_{eh} = \chi_{eh}(0) < 0$, we choose a finite
interaction strength $U$ and a very low (zero) temperature $T$ so that
we are very close to the singularity in vertex $\Gamma^{RPA}$ with $0
< 1 + U\chi_{eh} \ll 1$. We now rearrange the expansion beyond RPA in
such way that we remain all the time in the critical region of this
singularity. We then replace the interaction strength $U$ as an
expansion parameter by a new small scale $a = 1 + U\chi_{eh}$. We keep
this scale fixed during summation of diagrams and use it to separate
large (relevant) from small (irrelevant) quantities. We will discuss
ways how we can use the existence of a small (large) scale to simplify
the parquet equations. In fact we will see that it is not the scale
$a$ itself but rather $1/|\ln a|$ that is taken as a small parameter
in the expansion beyond RPA in the critical region of the
electron-hole vertex. By the parquet equations we effectively replace
the direct summation of diagrams beyond RPA and their renormalization
in the critical region of the singularity in the Bethe-Salpeter
equation from the electron-hole channel. In analogy with classical
criticality, RPA corresponds to a static mean-field approximation and
the parquet approach to renormalization-group dynamical corrections to
the mean-field critical behavior.

\subsection{Finite temperatures: mean-field
  approximation}\label{sec:Finite_temp}

By solving the parquet equations in the critical region of a
Bethe-Salpeter equation we assume that no other critical point emerges
beyond that found in the one-channel (mean-field) approximation. The
singularity in SIAM emerges in the electron-hole scattering channel
and hence the irreducible vertex from the electron-electron channel
$\Lambda^{ee}_{\uparrow\downarrow}(i\omega_n,i\omega_{n'},i\nu_m)$ is
singular in the variable $i\nu_m =0$. As no other critical scattering
appears in the model, the irreducible vertex from the electron-hole
channel
$\Lambda^{eh}_{\uparrow\downarrow}(i\omega_n,i\omega_{n'},i\nu_m)$
remains regular. We now neglect all finite differences in regular
functions and replace $\Lambda^{eh}$ with a constant, an effective
interaction $\overline{U}$. Inserting this ansatz in
Eq.~\eqref{eq:BS-eh} we obtain
\begin{subequations}\label{eq:Lambda-ee}
  \begin{equation}\label{eq:Lambda-ee-full}
    \Lambda^{ee}_{\uparrow\downarrow}(i\omega_n,i\omega_{n'}; i\nu_m) = U
    -\ \frac{\overline{U}^2\chi_{eh} (i\nu_m)}{1 + \overline{U}\chi_{eh}
      (i\nu_m)}\   .
  \end{equation}
  Being in the critical region where the denominator on the right-hand
  side of Eq.~\eqref {eq:Lambda-ee} is very small, we can neglect the
  contribution to $\Lambda^{ee}$ from the bare interaction and take
  into consideration only the low-frequency singular part of this
  vertex
  \begin{align}\label{eq:Lambda-ee-simplified}
    \Lambda^{sing}_{\uparrow\downarrow}(i\nu_m)& = -\
    \frac{\overline{U}^2\chi_{eh} (i\nu_m)}{1 + \overline{U}\chi_{eh}
      (i\nu_m)}\ .
  \end{align}
\end{subequations}
Neglecting finite differences due to the frequency dependence of the
regular vertex $\Lambda^{eh}$ we have won a simplified frequency
dependence of the singular vertex in form of the RPA vertex
$\Gamma^{RPA}$ from Eq.~\eqref{eq:RPA-vertex}. This simple form
enables us to control analytically the critical behavior of the
singular vertex $\Lambda^{ee}$.

We now insert the derived singular part of the irreducible vertex from
the electron-electron channel into Eq.~\eqref{eq:BS-ee} and obtain
\begin{multline}\label{eq:Lambda-eh-full}
  \overline{U} = U - \frac 1\beta \sum_{n''} \Lambda^{sing}(i\nu_{m +
    n' -n''}) \\ \times G_\uparrow(i\omega_{n''})
  G_\downarrow(i\omega_{n + n' + m - n''}) \\ \times
  \left[\overline{U} + \Lambda^{sing}(i\nu_{m + n - n''}) - U \right]\
  .
\end{multline}
This equation is evidently inconsistent, since its right-hand side is
frequency dependent. We neglected finite differences in the regular
electron-hole irreducible vertex on the left-hand side of
Eq.~\eqref{eq:Lambda-eh-full}.  We must do the same on the right-hand
side as well. It contains, however, singular functions. We hence
cannot demand equality of both sides for a specifically chosen
Matsubara frequency. To make our simplifying ansatz consistent we have
to first regularize the right-hand side of
Eq.~\eqref{eq:Lambda-eh-full} and grant equality only in a mean. To
achieve consistency we multiply both sides with
$G_\uparrow(i\omega_{n}) G_\downarrow(i\omega_{ m - n})$ and
$G_\uparrow(i\omega_{n'}) G_\downarrow(i\omega_{m - n'})$ and sum over
fermionic Matsubara frequencies $i\omega_n$ and $i\omega_{n'}$.  We
further put $m=0$ to keep the effective interaction real.  Since the
low-frequency contributions are the most relevant ones, we obtain
\begin{align} \label{eq:Lambda-eh-averaged} \overline{U}\chi_{ee} & =
  U\chi_{ee} -\frac{\left\langle G_\uparrow G_\downarrow
      L_{\uparrow\downarrow}^2\right\rangle} {\chi_{ee} +\left\langle
      G_\uparrow G_\downarrow L_{\uparrow\downarrow}\right\rangle}\ ,
\end{align}
where $\chi_{ee} = \beta^{-1} \sum_n G_\uparrow(i\omega_n)
G_\downarrow(i\omega_{ - n})$ is the static electron-electron bubble.
Further on we used a notation
\begin{subequations}\label{eq:Convolutions-def}
  \begin{equation}\label{eq:Convolutions-two}
    L_{\uparrow\downarrow}(i\omega_n) = \frac 1{\beta}
    \sum_{n'}G_\uparrow(i\omega_{n'}) G_\downarrow(i\omega_{-n'})
    \Lambda^{sing}_{\uparrow\downarrow}(i\nu_{-n -n'}) \ ,
  \end{equation}
  \begin{equation} \label{eq:Convolutions-three} \left\langle
      G_\uparrow G_\downarrow X\right\rangle = \frac 1{\beta}
    \sum_{n}G_\uparrow(i\omega_{n}) G_\downarrow(i\omega_{-n})
    X(i\omega_{n})\ .
  \end{equation} \end{subequations}
 
Equations~\eqref{eq:Lambda-ee}
and~\eqref{eq:Lambda-eh-averaged}-\eqref{eq:Convolutions-def} form a
closed set of relations determining the static effective interaction
$\overline{U}$ and the dynamical vertex $\Lambda^{sing}(i\nu_m)$ as
functionals of the one-particle propagators $G_\sigma$ and the bare
interaction $U$. They were derived and justified in the critical
region of a singularity in the Bethe-Salpeter equation from the
electron-hole channel. That is, the denominator on the r.h.s. of
Eq.~\eqref{eq:Lambda-ee} approaches zero. The form of the defining
equations for $\Lambda^{sing}(i\nu_m)$ and $\overline{U}$ enables
their application in the whole region of the input parameters. The
approximation remains consistent everywhere in the Fermi-liquid
regime.

\subsection{Zero temperature: low-frequency
  approximation}\label{sec:Zero_temp}

Separation of large scales (potentially divergent functions) and small
scales (regular functions) is incomplete in SIAM. To stay close to the
critical point of the singularity in $\Lambda^{sing}$ we need to know
the value of the effective interaction $\overline{U}$. We do not need
to know, however, the detailed frequency dependence of vertex
$\Lambda^{eh}$ but only its action in Bethe-Salpeter
equation~\eqref{eq:BS-eh}. With an effective static interaction we
make a replacement
\begin{multline*}
  \beta^{-1}\sum_{n"}\Lambda^{eh}_{\uparrow\downarrow}(i\omega_n,
  i\omega_{n"}; i\nu_m) G_\uparrow(i\omega_{n''})
  G_\downarrow(i\omega_{n''+ m})\\ \times
  \Lambda^{ee}_{\uparrow\downarrow}(i\omega_{n"}, i\omega_{n'};
  i\nu_m) \\ \to \overline{U} \chi_{eh}(i\nu_m)
  \Lambda^{ee}_{\uparrow\downarrow}(i\omega_{n}, i\omega_{n'}; i\nu_m)
\end{multline*}
in the exact parquet equation. This replacement is qualitatively
correct if we treat integrals (sums over Matsubara frequencies) of
singular functions as regular ones and as far as the effective
interaction $\overline{U}$ remains finite. We will see that in SIAM it
is always the case even in the limit $U\to \infty$.

Although the critical behavior of the singular vertex $\Lambda^{sing}$
does not depend on the value of the effective interaction
$\overline{U}$, we need its value to determine the asymptotic behavior
when the critical point is approached. We need to determine how fast
the small scale
\begin{equation}\label{eq:critical-scale}
  a = 1 + \overline{U}\chi_{eh} \to 0
\end{equation}
approaches zero as a function of the input parameters, in particular,
of the bare interaction strength $U$. It is evident that this result
depends on the function dependence $\overline{U}(U)$.

The averaging of Eq.~\eqref{eq:Lambda-eh-full} we used in the
preceding subsection to reach consistency in the determination of the
effective interaction is only one, mean-field way. We can proceed in a
more accurate way if we fully utilize the dominance of the
low-frequency asymptotics of the singular vertex. We can approximate
the singular vertex in the critical region with $a\ll 1$ as
\begin{equation}\label{eq:Lambda-critical}
  \Lambda^{sing}(\omega_+) \doteq \frac {\overline{U}}{a -
    i\overline{U} \chi' \omega/\pi}
\end{equation}
where we denoted $\chi' = i\pi\partial \chi_{eh}(\omega_+)
/\partial\omega|_{\omega=0}$ and $\omega_+ = \omega + i0^+$.  It is
clear that only frequencies of order $a\pi/\overline{U}\chi'$ are
relevant in integrals with this singular function. It means that only
frequencies of order $a/\overline{U}\chi'$ are relevant in regular
functions when integrated with the singular vertex. We hence replace
all values of the redundant frequencies in regular functions by zero.
This is possible only at zero temperature with a continuous
distribution of Matsubara frequencies. The nonsingular electron-hole
irreducible vertex $\Lambda^{eh}(i\omega_{n}, i\omega_{n'}; i\nu_m) =
\overline{U}$. Such a static replacement cannot be done, however, in
singular functions. Before we can set redundant frequencies to zero in
Eq.~\eqref{eq:Lambda-eh-full} we multiply both sides either by
$G_\uparrow(i\omega_{n}) G_\downarrow(i\omega_{ m - n})$ or by
$G_\uparrow(i\omega_{n'}) G_\downarrow(i\omega_{m - n'})$ and
integrate over $i\omega_n$ or $i\omega_{n'}$, respectively. The output
is a "sufficiently" regular function with maximally logarithmically
divergent terms proportional to $|\ln a|$ where we already can put all
external frequencies to zero. We then obtain another equation for the
effective interaction
\begin{align} \label{eq:Lambda-eh-low} \overline{U}\chi_{ee} & =
  U\chi_{ee} -\frac{L_{\uparrow\downarrow}^2} {1
    +L_{\uparrow\downarrow}}
\end{align}
where we denoted $L_{\uparrow\downarrow} = L_{\uparrow\downarrow}(0)$.
This approximation is applicable only at zero temperature with the
sums over Matsubara frequencies replaced by integrals. In this
approximation, however, the singularity in the two-particle vertex is
treated more accurately, since unlike the preceding construction we
treated here logarithmically divergent scales of order $|\ln a|$ more
sensitively.  Logarithmic terms may go lost or be washed out by
additional summation over fermionic Matsubara frequencies used in the
preceding subsection.  It means that our precision is determined by
the largest "regular" scale being now of order $|\ln a|$. There is no
such a small parameter in the construction with
Eq.~\eqref{eq:Lambda-eh-averaged}.

\section{Kondo scale and Kondo temperature}\label{sec:Kondo-scale}

We now proceed with solving the derived simplified parquet equations
for the singular vertex $\Lambda^{sing}(\omega_+)$ and the effective
interaction $\overline{U}$. We resort to the spin-symmetric case
$G_\uparrow = G_\downarrow$ and analyze the solution in the critical
region of the singularity with $a = 1 +\overline{U}\chi_{eh} = 1 -
\overline{U}\int_{-\infty}^\infty d\omega f(\omega)
\Im[G_+(\omega)^2]/\pi \to 0$.  We used an abbreviation $G_\pm(\omega)
\equiv G(\omega \pm i0^+)$. The new small dimensionless scale $a$
emerging in the critical region of the two-particle vertex is dominant
for the determination of the behavior of SIAM in the strong-coupling
regime. We call it a Kondo scale, since, as we will demonstrate below,
it gives origin to and controls the Kondo strong-coupling asymptotics
known from the exact Bethe-ansatz solution.

We first evaluate the dominant contributions from the singular vertex
$\Lambda^{sing}(\omega_+)$ to the integral $L_{\uparrow\downarrow}$ in
the limit $a\to 0$. We simultaneously use both reduction schemes,
Eq.~\eqref{eq:Lambda-eh-averaged} and Eq.~\eqref{eq:Lambda-eh-low}.
The former for non-zero temperatures and the latter at zero
temperature. Using the low-frequency asymptotics,
Eq.~\eqref{eq:Lambda-critical}, we obtain at finite temperatures
\begin{subequations}\label{eq:L}
  \begin{equation}\label{eq:L-full}
    L_{\uparrow\downarrow}(z) \doteq \frac {G(z)G(-z)}{\chi'}
    \int_0^\xi \frac {dx }{\tanh(bx/2)}\ \frac {x}{a^2 + x^2}
  \end{equation}
  where we denoted a dimensionless inverse temperature $b =
  \pi\beta/\overline{U}\chi'$, effective bandwidth $\xi =
  \overline{U}\chi'D/\pi $, and $\chi' = -4 \int_{-\infty}^{\infty} d
  x f(\Delta x + \overline{\mu}) x(x^2 + 1)^{-3} $. At low
  temperatures, at which the integral becomes singular and the
  asymptotic representation~\eqref{eq:L-full} holds, we further obtain
  \begin{multline}\label{eq:L-asymptotic}
    L_{\uparrow\downarrow}(z) \doteq \frac{ G(z)G(-z)}{2\chi'}\\
    \times \left[\ln\left( \frac{b^2}{1 + b^2a^2}\right) +
      \frac{4}{ba}\arctan\left( \frac 1{ba}\right) \right]\ .
  \end{multline}
  The leading singular contribution to the integral $L$ at zero
  temperature comes out rather simply
  \begin{align}\label{eq:L-zero}
    L_{\uparrow\downarrow}(z) \doteq \frac {G(z)G(-z)}{\pi^2\rho_0^2}\
    |\ln a | \
  \end{align}
\end{subequations}
where $\rho_0= \Delta/\pi(\Delta^2 + \overline{\mu}^2)$ is the density
of states at the Fermi energy.

Parquet equations~\eqref{eq:Lambda-ee}, \eqref{eq:Lambda-eh-averaged}
do not allow for a solution with a critical point, $a = 0$, at a
finite temperature, that is with with $ba=0$. The critical point can
only be asymptotically approached when the temperature approaches
zero. Actually, the critical behavior saturates when the product $ba$
becomes of order unity. This condition in the asymptotic limit
$b\to\infty$ defines a temperature at which the static two-particle
vertex function asymptotically approaches its maximal value. In the
Kondo regime, below this saturation temperature, called Kondo
temperature, physical quantities are saturated and the Kondo
temperature overtakes the role of the temperature scale. Using the
low-temperature asymptotics from Eq.~\eqref{eq:L-asymptotic} together
with the condition $ba=1$ in Eq.~\eqref{eq:Lambda-eh-averaged} we
obtain an expression for the Kondo temperature
\begin{subequations}\label{eq:Kondo}
  \begin{multline}\label{eq:Kondo-temperature}
    k_B T_K = \frac{\pi|\chi_{eh}|}{\chi'}\\ \times \exp \left\{ -
      \frac{\chi' \chi_{ee}}{|\chi_{eh}|}(U|\chi_{eh}| - 1) \frac
      {\langle (GG)_{ee}^2\rangle}{\langle (GG)_{ee}^3\rangle}\right\}
  \end{multline}
  where we denoted $\langle (GG)_{ee}^n\rangle = -\pi^{-1}
  \int_{-\infty}^\infty dx f(x)\Im\left[(G_+(x) G_-(-x))^n\right]$.
  Kondo regime sets in only asymptotically in the limit
  $U|\chi_{eh}(0)| \to \infty$. The genuine strong-coupling regime in
  SIAM sets in already for $U|\chi_{eh}(0)| > 1$. This regime is not
  reached by simpler weak-coupling approximations such as RPA or FLEX,
  where always $U|\chi_{eh}(0)| < 1$. Notice that the Kondo
  temperature cannot be defined unambiguously and
  Eq.~\eqref{eq:Kondo-temperature} is valid as order of magnitude
  only.

  Using Eq.~\eqref{eq:Lambda-eh-low} we cannot determine the Kondo
  temperature directly, since it is applicable only at zero
  temperature. We can nevertheless determine the Kondo scale $a$ as a
  function of the bare interaction strength. Inserting the asymptotic
  result from Eq.~\eqref{eq:L-zero} into Eq.~\eqref{eq:Lambda-eh-low}
  we obtain an equation for $|\ln a|$. From it we find a solution for
  the Kondo scale
  \begin{equation}\label{eq:Kondo-scale}
    a = \exp\left\{- \frac {\chi_{ee}}{|\chi_{eh}|} \left(U |\chi_{eh}| -
        1  \right) \right\}\ .
  \end{equation}
\end{subequations}

Expressions~\eqref{eq:Kondo-temperature} and~\eqref{eq:Kondo-scale}
for the Kondo temperature and the Kondo scale were derived for an
arbitrary form of the one-electron propagators. The only assumption
was that the electron-hole bubble $\chi_{eh}(z)$ is analytic at $z=0$.
We argued in Ref.~\onlinecite{Janis07a} that the Hartree one-electron
propagators deliver the best results in SIAM on both small and large
energy scales. Any \textit{dynamical} one-electron self-consistency
negatively interferes in the control of the two-particle singularity
in the Bethe-Salpeter equation won by the parquet approximation. In
particular, the electron-hole bubble $\chi_{eh}\to 0$ with
$U\to\infty$ when calculated with the fully renormalized one-electron
propagators, which forces the effective interaction to diverge. Our
simplifying scheme was justified for finite effective interaction
$\overline{U} <\infty$. The parquet approximation for the vertex
functions with $\mathcal{I} = U$ corresponds to a Hartree
approximation on the one-particle level.  Going beyond Hartree
propagators in the parquet scheme demands to renormalize adequately
the completely irreducible vertex $\mathcal{I}$ to keep the
approximation consistent. The only peremptory one-particle
self-consistency is the static one, where the effective chemical
potential $\overline{\mu}$ is calculated from the fully renormalized
one-electron propagator.

We hence use the Hartree propagators in our parquet approximation to
evaluate explicitly the two-particle functions in
Eqs.~\eqref{eq:Kondo}. In the low-temperature limit the dominant
contribution to the Kondo temperature comes form the zero-temperature
values of the functions on the right-hand side of
Eq.~\eqref{eq:Kondo-temperature}. At zero temperature the two-particle
bubbles read
\begin{subequations}\label{eq:chi-static}
  \begin{align}
    \chi_{eh} &= - \frac 1\pi \ \frac \Delta{\Delta^2 +
      \overline{\mu}^2}\ ,\\
    \chi_{ee}& = \frac 1{2\pi\overline{\mu}} \left[ \frac \pi 2 +
      \arctan \left(\frac{\overline{\mu}^2 -
          \Delta^2}{2\Delta\overline{\mu}} \right)\right] \ , \\
    \chi' &=\frac {\Delta^4}{(\Delta^2 + \overline{\mu}^2)^2}\ .
  \end{align}
\end{subequations}
Further on we obtain
\begin{subequations}
  \begin{align}\label{eq:GG}
    \langle (GG)_{ee}^2\rangle &= \frac 1{4\pi \overline{\mu}^3}
    \left[ \frac \pi 2 + \arctan \left(\frac{\overline{\mu}^2 -
          \Delta^2}{2\Delta\overline{\mu}}\right) - \frac {2\Delta
        \overline{\mu}}{\Delta^2 + \overline{\mu}^2}\right]\ ,\\
    \langle (GG)_{ee}^3\rangle &= \frac 1{16\pi \overline{\mu}^5}
    \left[ \frac {3\pi} 2 + 3 \arctan \left(\frac{\overline{\mu}^2
          -\Delta^2}{2\Delta\overline{\mu}}\right) \right. \nonumber
    \\ &\qquad \qquad\quad \left. -\ \frac {2\Delta
        \overline{\mu}(3\Delta^2 + 5 \overline{\mu}^2)}{(\Delta^2 +
        \overline{\mu}^2)^2}\right]\ .
  \end{align}
\end{subequations}
The effective interaction in the Kondo regime reads $\overline{U}= \pi
(1 - a) (\overline{\mu}^2 + \Delta^2)/\Delta$.

With the above representation we obtain an explicit dependence of the
Kondo temperature on the interaction strength $U$ and the effective
chemical potential $\overline{\mu}$. We explicitly mention only two
important limits. First, in the symmetric case $\overline{\mu}=0$ we
have
\begin{subequations}\label{eq:TK}
  \begin{equation}\label{eq:TK-symmetric}
    k_BT_K \doteq \Delta \exp\left\{ - \frac 53 [U\rho_0 -1]\right\}
  \end{equation}
  The Kondo temperature is determined by the Kondo dimensionless scale
  derived in Ref.~\onlinecite{Janis07a}.

  The other limit with a simple representation of the Kondo
  temperature is an almost empty impurity level where $\overline{\mu}
  \gg \Delta$.  In this case we obtain
  \begin{equation}\label{eq:TK-asymmetric}
    k_BT_K \doteq \frac{\overline{\mu}^2}{\Delta} \exp\left\{-  \frac
      {2\pi\Delta}{3\overline{\mu}}\left(
        \frac{U\Delta}{\pi\overline{\mu}^2} - 1\right) \right\}
  \end{equation}
\end{subequations}
where $\overline{\mu} = E_d + Un $. For $\overline{\mu}\gg\Delta$ we
explicitly have $\overline{\mu} = E_d(1 - \sqrt{1 + 4U\Delta/\pi
  E_d^2})/2$ and $n\sim \Delta/\pi\overline{\mu}$.  From which we get
$\overline{\mu} \sim |E_d|$ and $E_d^2 \sim U\Delta$.  We see that the
strong-coupling regime $U\Delta > \pi E_d^2$ sets in also for an
almost empty impurity level, but the Kondo temperature is no longer
exponentially small.

The Kondo scale calculated from Eq.~\eqref{eq:Kondo-scale} with
Hartree propagators reads
\begin {subequations}\label{Kondo-scale-explicit}
  \begin{multline}
    a = \exp\left\{- \frac{\pi/2 + \arctan(\frac{\overline{\mu}^2 -
          \Delta^2}{2\overline{\mu}\Delta})}{2\overline{\mu}\Delta}\right. \\
    \left. \times \left[\frac U\pi \Delta - \Delta^2 -
        \overline{\mu}^2 \right] \right\}\ .
  \end{multline}
  In the symmetric case and for an almost empty impurity level it
  reduces to
  \begin{align}\label{eq:a-symmetric}
    a &= \exp\{-(U\rho_0 - 1)\}
  \end{align}
  and
  \begin{align}
    \label{eq:a-empty} a &= \exp\left\{- \frac{\pi}{2
        |E_d|\Delta}\left(U\Delta - E_d^2 \right) \right\}\,
  \end{align}
\end{subequations}
respectively. Using further the solution for $E_d$ in the limit $n\to
0$ we obtain an asymptotic dependence $a\sim
\exp\left\{-\frac{\alpha\pi}{2\Delta}|E_d|\right\}$ or $a\sim
\exp\left\{-\alpha'\sqrt{\frac{U}{\Delta}}\right\}$ for $U\to \infty$.
The Kondo regime sets in for almost empty impurity level if $Un^2 >
\pi^2\Delta$.

Both expressions~\eqref{eq:TK-symmetric} and~\eqref{eq:a-symmetric}
for the Kondo temperature in the symmetric case qualitatively agree
(in universal features) with the exact formula $k_BT_K \doteq
\sqrt{U\Delta/2}\exp\{\pi E_d (E_d + U)/2\Delta U \}$ known from the
$s-d$ exchange Hamiltonian asymptotically reached in the limit
$U\to\infty$ of SIAM.\cite{Tsvelik82} The Kondo temperature for an
almost empty impurity level, Eq.~\eqref{eq:TK-asymmetric}, deviates in
the dependence on $E_d$ from the exact result. The two asymptotic
results were, however, derived under different conditions.  The former
representation was obtained for a fixed $\overline{\mu} \gg \Delta$
while the latter for a fixed $E_d$ constrained to $E_d \ll -\Delta$
and $E_d + U \gg \Delta$.\cite{Haldane78} The Kondo scale from
Eq.~\eqref{eq:a-empty} improves upon this imperfection in the
asymmetric situation of the approximate scheme based on
Eq.~\eqref{eq:Lambda-eh-averaged} . It qualitatively well reproduces
the exact result with the exponent linearly proportional to the
position of the impurity level $E_d$. As expected, the
zero-temperature, low-frequency approximation on the effective
interaction $\overline{U}$, Eq.~\eqref{eq:Lambda-eh-low}, better
reproduces the dependence of the Kondo scale on the interaction
strength and in particular on the position of the impurity level and
the impurity-level occupation.

\section{Self-energy: Kondo resonance and Hubbard satellite bands}
\label{sec:Self-energy}

\begin{figure}
  \includegraphics[scale=0.8]{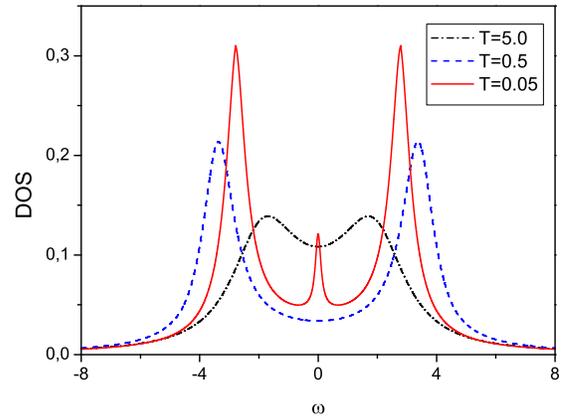} \caption{Density of
    states for $U=4$ and various temperatures demonstrating formation
    of the Kondo resonance at the Fermi energy for the symmetric case.
    The zero-temperature value of DOS in the half-filled case due to
    the Friedel sum rule is $1/\pi$.} \label{fig:fig1}
\end{figure}
\begin{figure}
  \includegraphics[scale=0.8]{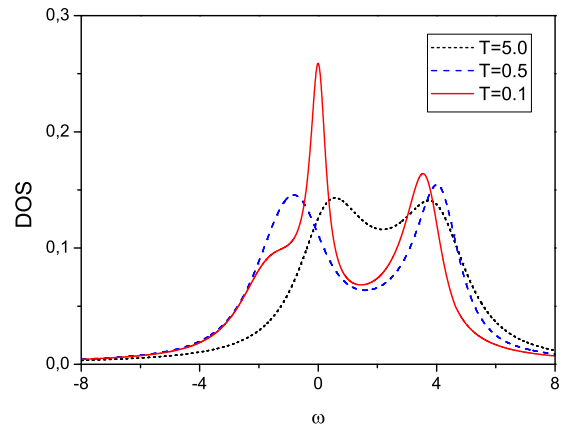} \caption{Temperature
    dependence of the density of states for a partial occupation of
    the impurity level $n=0.4$ and $U=4$.} \label{fig:fig2}
\end{figure}

We defined the Kondo temperature and the Kondo scale from a critical
behavior of the two-particle vertex near a singularity in the
electron-hole correlation function.  The one-electron propagators were
used as input parameters for the two-particle vertices.  In this
situation with one-electron propagators chosen independently of the
two-particle ones we cannot interpret the two-particle functions we
use as the actual physical quantities they diagrammatically stand for.
We have to conform with conservation laws in consistent theories. To
make the approximation thermodynamically consistent, we have to
introduce the self-energy as a functional of the two-particle vertex.
The physical quantities in conserving theories are then derived from
the self-energy via appropriate (functional) derivatives.

We use the Schwinger-Dyson equation to relate the self-energy with the
two-particle vertex.\cite{Janis07a} In our case of the simplified
parquet equations it reduces to
\begin{equation}
  \label{eq:Parquet-Sigma} \Sigma_\sigma(i\omega_n)=\frac{U}{\beta
  }\sum_{n'} \frac {G_{-\sigma}(i\omega_{n'}) }{1 +
    \overline{U}\chi_{eh}(i\omega_{n'} - i\omega_{ n})}\ .
\end{equation}
In the spin symmetric case and in the Kondo regime (critical region of
the singularity in the two-particle vertex) we again use the
low-frequency expansion of the denominator in the analytically
continued version of Eq.~\eqref{eq:Parquet-Sigma}. We further employ
the electron-hole symmetry of the Hartree propagators $G( i\omega_n -
\bar{\mu}) = - G(-i\omega_n + \bar{\mu})$ to guarantee analytic
properties also in approximate numerical evaluations of the
self-energy.  After analytic continuation we explicitly obtain for the
real and imaginary parts of the self-energy
\begin{subequations} \label{eq:Sigma-Kondo}
  \begin{align}
    \Re\Sigma(y) &= \frac{U|\chi_{eh}(0)|}{2\pi(1 - a) \chi'}
    \int_{-\infty}^{\infty}\frac {dx}{(x + y - \overline{\mu})^2 + 1}
    \ \frac {1 }{x^2 + \widetilde{a}^2} \nonumber \\ &\qquad \times
    \left[\frac{x(x+y-\overline{\mu})}{\tanh(\beta x/2)} -
      \widetilde{a} \tanh\left(\beta(x+y)/2 \right) \right] , \\ %
    \Im\Sigma(y) &= - \frac{U|\chi_{eh}(0)|}{2\pi(1 - a) \chi'}
    \int_{-\infty}^{\infty}\frac {dx}{(x + y - \overline{\mu})^2 + 1}
    \ \frac {x }{x^2 + \widetilde{a}^2}\nonumber \\ & \qquad\qquad
    \times \left[\frac{1}{\tanh(\beta x/2)} - \tanh\left(\beta (x+y)/2
      \right) \right] ,
  \end{align}\end{subequations}
where we set $\Delta =1$ and denoted $\widetilde{a} = a \pi
|\chi_{eh}(0)|/(1 - a)\chi'$. Representation~\eqref{eq:Sigma-Kondo} is
meaningful for $0\le a \le 1$, but it is justified in the asymptotic
limit $\beta\to\infty$ and $\beta a \sim 1$. The physics in the Kondo
regime of SIAM, including the effective chemical potential
$\overline{\mu}$, in this approximation is determined from
self-energy~\eqref{eq:Sigma-Kondo} and its dependence on external
sources entering the theory via the Hartree propagators. For a
numerical solution, not deep in the Kondo regime, it is
straightforward to iterate Eq.~\eqref{eq:Lambda-ee} with either
Eq.~\eqref{eq:Lambda-eh-averaged} or Eq.~\eqref{eq:Lambda-eh-low} and
use their solution in Eq.~\eqref{eq:Parquet-Sigma} to determine
dynamical and thermodynamic properties of SIAM.

\begin{figure}
  \includegraphics[scale=0.8]{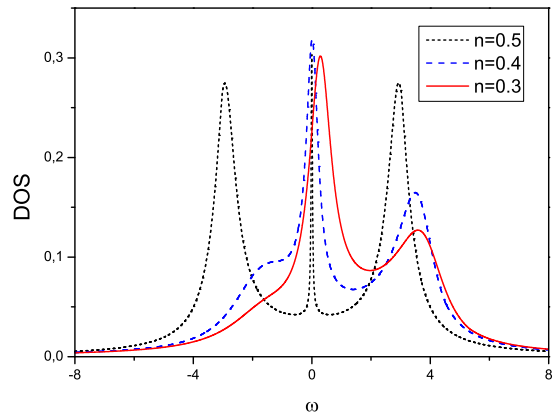}
  \caption{Dependence of the Kondo resonant peak on filling of the
    impurity level for $U=4$. Merging of the quasiparticle peak with
    the lower Hubbard band for lower fillings is apparent.}
  \label{fig:fig3}
\end{figure}

Figure~\ref{fig:fig1} shows temperature dependence of the spectral
density for $U=4$ and half filling. When the temperature decreases,
quasiparticle states are first expelled from the Fermi energy to
satellite peaks.  But at temperatures of order of the Kondo one, a new
resonance around the Fermi energy starts to develop so that a
metal-insulator transition is screened. This scenario takes place at
arbitrary filling, but the exponentially narrow Kondo resonance
realizes practically only for fillings close to the symmetric case,
Fig.~\ref{fig:fig2}. Farther apart from the half-filled case the Kondo
temperature increases, the quasiparticle peak broadens and gradually
merges with the lower satellite band, Fig.~\ref{fig:fig3}.  Moreover,
the weight of the Kondo peak also decreases with doping the symmetric
case with either holes or particles. The numerical results are in a
good agreement with the behavior obtained by NRG in
Ref.~\onlinecite{Costi94}. It is important that the simplified parquet
approximation used here predicts qualitatively correctly formation of
a Kondo peak near the Fermi energy also in slightly asymmetric
situations where the Friedel sum rule (obeyed by this approximation)
pinning the zero-temperature density of states at the Fermi energy no
longer holds. The Kondo peak in asymmetric fillings is hence a
consequence of genuine strong electron correlations. Only theories
producing a Kondo resonance peak in the asymmetric case can be relied
upon in the strong-coupling regime.

\begin{figure}
  \includegraphics[scale=0.8]{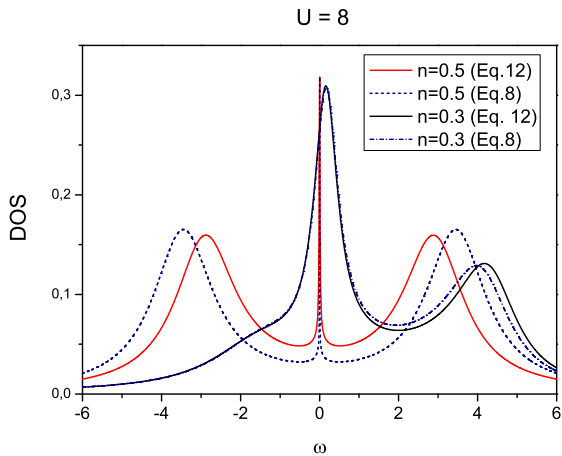} \caption{Spectral
    function calculated with the effective interaction calculated from
    Eq.~\eqref{eq:Lambda-eh-averaged} and Eq.~\eqref{eq:Lambda-eh-low}
    for $U=8$. The two approximate schemes do not significantly differ
    at any filling except for a narrow asymptotic critical region of
    asymmetric models.} \label{fig:fig4}
\end{figure}

We used Eq.~\eqref{eq:Lambda-eh-averaged} for the determination of the
effective interaction $\overline{U}$ in the calculations of the
self-energy and the spectral function. The differences in the results
obtained with the other approximate decoupling of the parquet
equations in Eq.~\eqref{eq:Lambda-eh-low} are not dramatic. The
results are qualitatively the same with minor quantitative deviations
as demonstrated in Fig.~\ref{fig:fig4}. In the symmetric situation the
distance between the satellite peaks is smaller and the central peak
is slightly broader when Eq.~\eqref{eq:Lambda-eh-low} is used.
 
\begin{figure}
  \includegraphics[scale=0.8]{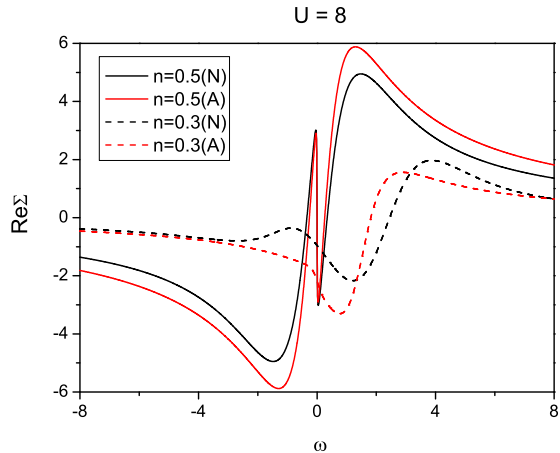} \caption{Real part of the
    self energy from the numerical solution (N) and the asymptotic
    form from Eq.~\eqref{eq:SE-asymptotic} (A) at zero temperature and
    two fillings. The critical behavior is more pronounced in the
    asymptotic solution} \label{fig:fig5}
\end{figure}

Critical behavior of the two-particle vertex in the Kondo regime
($a\ll 1$) allows one to assess the self-energy and the resulting
spectral function analytically. Using the fact that $a\to 0$ in
Eqs.~\eqref{eq:Sigma-Kondo} we obtain for the leading asymptotic
behavior of the self-energy at zero temperature
\begin{subequations}\label{eq:SE-asymptotic}
  \begin{multline}\label{eq:SE-real}
    \Re\Sigma(\omega_+) = \frac{U\ln\left[1 +
        \frac{\overline{U}^2\pi^2 \rho_0^4 \xi^2}{a^2}
      \right]}{2\overline{U}\pi^2 \rho_0^2}\ \Re \mathcal{G}(\omega_+)
    \\ + \frac{U\arctan\left(\frac{\overline{U}\pi \rho_0^2\omega}a
      \right)}{\overline{U}\pi^2\rho_0^2}\ \Im \mathcal{G}(\omega_+)\
    ,
  \end{multline}
  \begin{equation}\label{eq:SE-imaginary}
    \Im\Sigma(\omega_+) = \frac{U\ln\left[1 +
        \frac{\overline{U}^2\pi^2 \rho_0^4 \omega^2}{a^2}
      \right]}{2\overline{U}\pi^2 \rho_0^2}  \ \Im \mathcal{G}(\omega_+)\ .
  \end{equation}
\end{subequations}
Representation~\eqref{eq:SE-asymptotic} is quite general and holds in
the Kondo regime for any form of the one-electron propagator
$\mathcal{G}$ used in the parquet approximation. That is, we can
analyze with this representation any dispersion relation or
approximate form of the one-electron propagators used in the
Schwinger-Dyson equation~\eqref{eq:Parquet-Sigma}. Most importantly,
we can assess the width of the Kondo resonance near the Fermi energy
and decide about the existence of the satellite upper and lower
Hubbard bands. In Fig.~\ref{fig:fig5} we compared the real-part of the
self-energy calculated with the full representation,
Eq.~\eqref{eq:Sigma-Kondo}, and the asymptotic form,
Eq.~\eqref{eq:SE-asymptotic}. We can see that there are no relevant
(qualitative) differences in the results and the agreement is almost
perfect in the critical region and for small energies near the Fermi
energy. The asymptotic formula should hold in the critical region with
$a\ll 1$ and hence the deeper in the critical (Kondo) region we are
the better the agreement becomes.

We already proved in Ref.~\onlinecite{Janis07a} that one-particle
self-consistent approximations with fully renormalized one-electron
propagators smear out the satellite Hubbard bands. On the other hand,
numerical calculations clearly demonstrate that the approximation with
the Hartree one-electron propagators leads to satellite bands. With
the asymptotic formulas~\eqref{eq:SE-asymptotic} we can even prove it
and determine the positions of these bands.
 
We use Hartree propagators in Eqs.~\eqref{eq:SE-asymptotic}.  The
effective chemical potential $\overline{\mu}$ is determined from the
actual occupation of the impurity level. We first estimate the
position of the Kondo resonance peak that is, maximum of the spectral
function very close to the Fermi energy, within a frequency region of
order of the Kondo scale. We determine the center of the Kondo peak
from vanishing of the imaginary part of the one-electron propagator.
Maximum of the spectral function at the Fermi energy is guaranteed by
the Friedel sum rule only at half filling. Using the Hartree
propagators on the right-hand side of Eqs.~\eqref{eq:SE-asymptotic}
for the self-energy we obtain a solution for $\partial
\Im\left[\omega_+ - \overline{\mu} -
  \Sigma(\omega_+)\right]^{-1}/\partial\omega = 0$ in a form
\begin{multline}\label{eq:X-imaginary}
  \tilde{x} = \\ \frac{2 U (\pi + U \ln\sqrt{1 + \tilde{x}^2}\ )
    (\arctan(\tilde{x}) + \overline{\mu}|\ln a |)}
  {U^2(\arctan(\tilde{x}) + \overline{\mu}|\ln a|)^2 - (\pi +
    U\ln\sqrt{1 + \tilde{x}^2}\ )^2 }
\end{multline}
where we denoted $\tilde{x}= \omega/a(1 + \overline{\mu}^2)$ and
assumed that it is of order unity. In the symmetric case,
$\overline{\mu}=0$, we have a trivial solution $x=0$. The position of
the Kondo resonance peak moves off the Fermi energy in the opposite
direction given by the effective chemical potential $\overline{\mu} =
E_d + Un$. For small values of $\overline{\mu}$ we explicitly obtain
\begin{align}\label{eq:X-small-mu}
  \omega_0 & = -\ \frac{2 U\overline{\mu}\ a |\ln a|}{\pi}
\end{align}
Maximum of the spectral function for larger effective chemical
potentials $\overline{\mu}\gg \pi/|\ln a |$ eventually moves out from
a region of order of the Kondo scale around the Fermi energy.

The width of the Kondo resonance can be estimated from the distance
between the opposite points of steepest descent of the spectral
function around the peak near the Fermi energy. They can be obtained
as points where the derivative of the real part of the one-electron
propagator vanishes. It happens for
\begin{multline}\label{eq:X-real}
  \tilde{x} = \\ \frac{(\pi + U\ln\sqrt{1 + \tilde{x}^2}\ )^2 -
    U^2(\arctan(\tilde{x}) + \overline{\mu}|\ln a|)^2 }{2
    U(\arctan(\tilde{x}) + \overline{\mu}|\ln a|) (\pi + U \ln\sqrt{1
      + \tilde{x}^2}\ )}
\end{multline}
For small effective chemical potentials, $\overline{\mu}\ll \pi/2| \ln
a|$, Eq.~\eqref{eq:X-real} has two solutions $\tilde{x}\sim 1$. The
Kondo peak is then well formed. It broadens with increasing the
effective chemical potential.  Above the critical value
$\overline{\mu} = \pi/2|\ln a|$ one of the solutions is pushed to
(minus) infinity and the Kondo peak dissolves in the bulk of other
states.

The positions of the centers of the satellite Hubbard bands can be
determined from vanishing of the imaginary part of the full
one-particle propagator. We have two solutions apart from the Fermi
energy. The centers of the satellite peaks are determined from a
solution proportional to the logarithm of the Kondo scale $|\ln a|$.
We obtain explicitly for the Hartree propagators in SIAM
\begin{equation}\label{eq:Hubbard-bands}
  \Omega_\pm = \overline{\mu} \pm \sqrt{\frac{U|\ln
      a|}{\overline{U}\pi^2\rho_0^2}} = \overline{\mu} \pm \frac
  {\sqrt{U\chi_{ee}(U|\chi_{eh}| - 1)}}{\pi|\chi_{eh}|}
\end{equation}
where the two-particle bubbles $\chi_{ee}$ and $\chi_{eh}$ are
determined from Eqs.~\eqref{eq:chi-static}. In the symmetric case we
have $\Omega_\pm = \pm U/\pi$. This result deviates from the atomic
limit levels $\Omega_\pm = \pm U/2$, but we must realize that the
Hubbard satellite peaks in the Kondo regime are not related to the
atomic limit, since the latter is not (cannot be) part of the solution
with a singular two-particle vertex. Beware that neither the exact
Bethe-ansatz solution incorporates the atomic limit. Away from the
half-filled impurity level, the distance between the satellite peaks
increases with the square root of the effective chemical potential
$\overline{\mu}$. Eq.~\eqref{eq:Hubbard-bands} also tells us that the
Hubbard satellite bands may exist only in the strong-coupling regime
defined as $U|\chi_{eh}| > 1$. The Kondo regime sets in only later for
$U|\chi_{eh}| \gg 1$.

\section{Conclusions}\label{sec:Conclusions}

We studied the single-impurity Anderson model in the strong-coupling
regime with the aim to develop a reliable and analytically
controllable approximation in this regime. We succeeded in simplifying
the parquet equations to a manageable form without loosing consistency
and fundamental characteristics of the Fermi-liquid regime for
arbitrary fillings of the impurity level. We showed that the derived
simplified parquet equations for two-particle irreducible vertices in
the electron-hole ($\Lambda^{eh}$) and electron-electron
($\Lambda^{ee}$) singlet channels result from a resummation of the
diagrammatic perturbation expansion in the critical region of a
singularity in the two-particle vertex.  Such a singularity emerges in
SIAM due to multiple electron-hole scatterings. The resummation of the
perturbation expansion proceeds so that the critical region is reached
first within RPA. Diagrammatic contributions beyond RPA are then
summed not to be driven out of the critical region of the two-particle
vertex. Singularity in the two-particle vertex is controlled by a
vanishing scale. This new scale replaces the interaction strength in
the perturbation expansion beyond RPA and is used to select relevant
contributions in the critical regime.

The principal assumption for rearranging the diagrammatic expansion in
a two-particle criticality is to explicitly sum and control
two-particle contributions. We distinguish two basic classes of
two-particle functions: singular and regular. The former diverge with
vanishing of the critical scale and the latter remain finite.
Logarithmically divergent functions (integrals over the singular
functions) are treated as (marginally) regular ones. The critical
scale enables us to neglect finite differences in the regular
two-particle functions and replace them with constants. Only the
potentially divergent functions are kept dynamical in the two-particle
criticality. Even more, only the singular low-frequency asymptotics
matters in the critical region.

It is clear that contributions beyond RPA must be summed in a
self-consistent way to make the perturbation theory of SIAM
singularity-free for any finite interaction strength.  The parquet
equations provide us with the necessary two-particle self-consistency.
We presented two ways how to simplify the parquet equations in the
critical region of the singularity in the electron-hole channel. They
differ in the way we suppress the frequency dependence of regular
functions. This ambiguity, however, influences only non-universal
critical properties. In the first construction we replaced the regular
vertices with their specific averaged values. In the second one we
used the fact that only frequencies of order of the vanishing critical
scale remain important. In both cases we reached a set of manageable
equations determining an effective interaction, replacing the bare one
in RPA, and the singular two-particle vertex.
 
The solution of the simplified parquet equations led to the Kondo
behavior in the strong-coupling regime for all fillings of the
impurity level. We found that the vanishing critical scale of the
two-particle vertex is the one determining the Kondo temperature at
which the two-particle vertex saturates and the Kondo temperature
overtakes the control of the low-temperature behavior of two-particle
vertex and correlation functions. We thus find that the Kondo behavior
is just a critical behavior due to a singularity in the two-particle
vertex induced by frequency fluctuations. The critical point in SIAM
is reached only by infinite bare interaction, but the critical region
extends deep into the physical domain of its finite values. Unlike the
existing (approximate) solutions of SIAM we clearly demonstrated the
two-particle origin of the Kondo behavior. In fact, it is not the
density of states on the Fermi surface $\rho_0$ that controls the
Kondo behavior but rather the electron-hole bubble $|\chi_{eh}|$.  It
is fortunate that for the Lorentzian DOS used here and in the Bethe
ansatz the two numbers coincide.
 
Finally we investigated one-electron functions in the Kondo regime. We
related the two-particle Kondo scale with the width of the
quasiparticle peak near the Fermi energy and found a good qualitative
agreement with the exact result. We were able to estimate analytically
the width of this peak and showed that it is well formed only close to
the half-filled case. Farther away from the symmetric situation the
Kondo resonance dissolves in the bulk of other states.  We also showed
that our parquet approximation not only reproduces correctly the Kondo
resonance in the spectral function but it is also able to produce the
satellite Hubbard bands. We found a criterion for their existence and
an estimate for the positions of their centers.

To summarize, we presented a simple analytic approximation producing
dynamical properties of SIAM for all interaction strengths and
impurity fillings. It correctly reproduces universal features of the
low-temperature Kondo behavior and predicts a Kondo resonance peak
near the Fermi energy in the strong-coupling regime for small
deviations from the electron-hole symmetric case. At intermediate
couplings it is in a good quantitative agreement with more laborious
numerical renormalization-group calculations. We identified the Kondo
behavior as the critical behavior near a singularity in the
electron-hole Bethe-Salpeter equation with balanced electron-hole and
electron-electron multiple scatterings. The electron-hole scatterings
are needed to reach a singularity in Bethe-Salpeter equations and the
electron-electron ones to screen the inter-particle interaction. Due
to its universality, the approximation can easily be extended to more
complex multi-orbital or translationally invariant lattice models for
a reliable qualitative and quantitative investigation of a transition
from weak to strong-coupling regimes in materials with tangible
electron correlations.

\acknowledgments \noindent This research was carried out within
project AVOZ10100520 of the Academy of Sciences of the Czech Republic
and supported in part by Grant No.  202/07/0644 of the Grant Agency of
the Czech Republic. We thank Matou\v s Ringel for valuable and
inspiring discussions.

\end{document}